# The Wide Field Monitor onboard the eXTP mission

M. Hernanz[*,a1,a2], S. Brandt[b], M. Feroci[c1,c2], P. Orleanski[d], A. Santangelo[f],
S. Schanne[g], Xin Wu[e], J. in't Zand[h], S.N. Zhang[i], Y.P. Xu[i], E. Bozzo[j], Y. Evangelista[c1,c2], J. L. Gálvez[a1,a2],
C. Tenzer[f], F. Zwart[h], F.J. Lu[i], S. Zhang[i], T.X. Chen[i],
F. Ambrosino[c1], A. Argan[c1,c2], E. Del Monte[c1,c2], C. Budtz-Jørgensen[b], N. Lund[b], P. Olsen[b], C. Mansanet[a1],
R. Campana[k], F. Fuschino[k], C. Labanti[k], A. Rachevski[l], A. Vacchi[l,m], G. Zampa[l], N. Zampa[l], I.
Rashevskaya[n], P. Bellutti[o], G. Borghi[o], F. Ficorella[o], A. Picciotto[o], N. Zorzi[o],
O. Limousin[g], A. Meuris[g],
on behalf of the eXTP Consortium[p]

[a1]Institute of Space Sciences (ICE, CSIC), Campus UAB, C/ Can Magrans s/n, 08193 Cerdanyola del Vallès (Barcelona), Spain,
[a2]Institut d'Estudis Espacials de Catalunya (IEEC), 08034 Barcelona, Spain,
[b]DTU Space - National Space Institute, Elektrovej Building 327, DK-2800 Kgs. Lyngby, Denmark,
[c1]INAF/IAPS, Via Fosso del Cavaliere 100, I-00133 Roma, Italy,
[c2]INFN – Roma Tor Vergata, Via della Ricerca Scientifica 1, I-00133 Roma, Italy,
[d]Space Research Center, Polish Academy of Sciences, Bartycka 18a, PL-00-716 Warszawa, Poland,
[e]DPNC, Geneve University, Quai Ernest-Ansermet 30, 1205 Geneva, Switzerland,
[f]Institut für Astronomie und Astrophysik, Eberhard Karls Universität, 72076 Tübingen, Germany,
[g]CEA Paris-Saclay/IRFU, F-91191 Gif sur Yvette, France,
[h]SRON Netherlands Institute for Space Research, Sorbonnelaan 2, 3584 CA Utrecht, the Netherlands,
[i]Key Laboratory for Particle Astrophysics, Institute of High Energy Physics, Beijing 100049, China
[j]Department of Astronomy, University of Geneva, Chemin d'Ecogia 16, 1290, Versoix, Switzerland,
[k]INAF/OAS, Bologna, Italy, [l]INFN Sez. Trieste, Trieste, Italy, [m]Univ. of Udine, Udine, Italy,
[n]INFN – TIFPA, Trento, Italy, [o]FBK, Trento, Italy
https://www.isdc.unige.ch/extp/[p]

## ABSTRACT

The eXTP (enhanced X-ray Timing and Polarimetry) mission is a major project of the Chinese Academy of Sciences (CAS) and China National Space Administration (CNSA) currently performing an extended phase A study and proposed for a launch by 2025 in a low-earth orbit. The eXTP scientific payload envisages a suite of instruments (Spectroscopy Focusing Array, Polarimetry Focusing Array, Large Area Detector and Wide Field Monitor) offering unprecedented simultaneous wide-band X-ray timing and polarimetry sensitivity. A large European consortium is contributing to the eXTP study and it is expected to provide key hardware elements, including a Wide Field Monitor (WFM). The WFM instrument for eXTP is based on the design originally proposed for the LOFT mission within the ESA context. The eXTP/WFM envisages a wide field X-ray monitor system in the 2-50 keV energy range, achieved through the technology of the large-area Silicon Drift Detectors. The WFM will consist of 3 pairs of coded mask cameras with a total combined Field of View (FoV) of 90x180 degrees at zero response and a source localization accuracy of ~1 arcmin. In this paper we provide an overview of the WFM instrument design, including new elements with respect to the earlier LOFT configuration, and anticipated performance.

---

[*] hernanz@ice.csic.es, phone: +34 937379788, www.ice.csic.es

# 1. INTRODUCTION

The eXTP is a science space mission project aimed to study matter under extreme conditions of density, gravity and magnetism[2]. It is a major project of CAS (Chinese Academy of Sciences) and CNSA (China National Space Administration). eXTP includes a large European contribution, based on the heritage from LOFT (Large Observatory for x-ray Timing), one of the M3 missions selected by ESA in 2011 for feasibility study[8,9,12,13]. eXTP has recently started an extended Phase A funded by China (during 2018) and its expected launch date is 2025.

The scientific payload of eXTP includes four instruments[27,28] the SFA (Spectroscopic Focusing Array), the LAD (Large Area Detector), the PFA (Polarimeter Focusing Array) and the WFM (Wide Field Monitor). The energies covered by the four instruments range from 0.5 to 50 keV, with very good spectral resolution and excellent timing capabilities. An artist's impression of the eXTP satellite is shown in Fig. 1 (see as well the paper about eXTP instrumentation, Santangelo et al., submitted to Science China, 2018).

The SFA is an array of nine identical X-ray telescopes, working in the energy range (0.5-10) keV, with energy resolution better than 180 eV (FWHM) at 6 keV. The SFA angular resolution is required to be less than 1 arcmin (HPD). In the current baseline, the SFA focal plane detectors are silicon-drift detectors (SDDs), that combine CCD-like spectral resolutions with very small dead times, and therefore are excellently suited for studies of the brightest cosmic X-ray sources at the smallest time scales. The effective area expected is about 7000 $cm^2$ at 2 keV and 5000 $cm^2$ at 6 keV[5,27].

The LAD instrument for eXTP is based on the LAD design for LOFT, but with a reduced number of detector modules[14,15,26,28]. It has a modular design, with 40 modules, each hosting 4x4 large area SDDs (a total of 640 units) and 4x4 capillary plate collimators. The LAD is a collimated narrow FoV instrument, 1 degree (FWHM), with a large collecting area, 3.4 $m^2$ at 8 keV, obtained with non-imaging SDDs, in the energy range (2-30) keV, with a spectral resolution better than 240 eV at 6 keV. The LAD and the SFA together reach an unprecedented effective area larger than 4 $m^2$

The PFA includes four identical X-ray focusing telescopes with focal plane imaging detectors - Gas Pixel Detectors (GPDs) able to determine the polarization of the incident X-rays[2]. The angular is better than about 30 arcsec (HPD). The energy band is (2-10) keV, with energy resolution typical of gas detectors - 15-20% at 6 keV - and effective area > 900 $cm^2$ at 2 keV. It reaches a minimum detectable polarization (MDP) of 5% in 100 ks for a source with a Crab-like spectrum of flux $3 \times 10^{-11}$ $erg.s^{-1}.cm^{-2}$, i.e., about 1 mCrab[11,27].

The WFM instrument for eXTP is based on the WFM design for LOFT[3], but with a slightly reduced number of cameras. It includes six coded mask cameras, keeping the main capabilities and performance of the LOFT/WFM. In this paper a detailed description of the eXTP/WFM is presented.

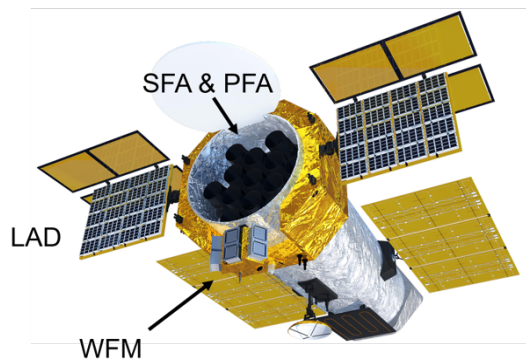

Figure 1. Artist's impression of the e-XTP spacecraft, with its science payload consisting of four instruments: the Spectroscopic and Polarimetric Focusing Arrays SFA and PFA, the Large Area Detector LAD and the Wide Field Monitor WFM.

# 2. SCIENCE OBJECTIVES AND REQUIREMENTS

The main purpose of the WFM is to detect sources for follow-up observations with the LAD and the other pointing instruments; therefore, the field of view is designed to have a maximum overlap with the sky accessible to LAD pointings. These sources are new transients, as well as known sources undergoing spectral state changes.

The primary goal of the WFM is to provide triggers for the SFA, LAD and SFA, with less than one day reaction time. Some examples follow:

- Detection of new, rare transient X-ray sources with ~1arcmin accuracy. For instance, black hole transients (core science topic: Strong Field Gravity).
- Detection of recurrent transient X-ray sources, primarily neutron stars (core science topic: Dense Matter - Equation of State).
- Detection of state changes in persistent X-ray sources, e.g., neutron stars and black holes (core science).

A FoV as wide as possible is required to catch such rare events. It is worth noticing that except for the new unknown transients, the source positions are known to roughly 1 arcsec resolution, thanks to previous observations in X-rays (with XMM-Newton, Chandra, Swift) and at other wavelengths.

Other goals of the WFM are related to what can be called "its own science" or more generally, "Observatory Science". Some examples:

- Imaging of the LAD FoV to determine if there is source confusion/contamination.
- Monitor the long-term behavior of X-ray sources.
- Detect short (0.1-100 s) bursts and transient events, recording data with full resolution
- Option to transmit the position of burst sources to ground in real time: adaptation of the "LBAS – LOFT Burst Alert System" to eXTP. In fact, LBAS was inspired on the SVOM Sino-French near future mission[23].

For a detailed description of the WFM observatory science goals, see the LOFT White papers[29,40] and the "Observatory Science with eXTP" paper (in't Zand et al., submitted to Science China, 2018).

A summary of the main scientific requirements of the WFM[3,8] are listed in Table 1. They have been derived from the whole set of WFM scientific goals; the corresponding anticipated performances are also shown. These are based on the work on LOFT/WFM simulations[6] and on LOFT background studies[4], among others.

Table 1. Summary of the WFM scientific requirements and anticipated performances

| Item | Requirement | Anticipated performance |
|---|---|---|
| Location accuracy | < 1 arcmin | < 1 arcmin |
| Angular resolution (FWHM) | < 5 arcmin | < 4.3 arcmin |
| Peak sensitivity in LAD direction (5$\sigma$) | 1 Crab (1s)<br>5 mCrab (50 ks) | 0.6 Crab (1s)<br>2.1mCrab (50 ks) |
| Absolute flux accuracy | 20% | <20% |
| Field of view | 1 $\pi$ = 3.1 steradian around the LAD pointing | 1.75 $\pi$ = 5.5 steradian at zero response<br>1.33 $\pi$ = 4.2 steradian at 20% of peak camera response |
| Energy range | 2–50 keV | 2–50 keV |
| Energy resolution (FWHM) | 500 eV (FWHM) @ 6 keV | < 300 eV @ 6 keV |
| Energy scale accuracy | 4% | < 2% |
| Energy bands for compressed images | 64 | ≥ 64 |
| Time resolution | 300 s for images<br>10 $\mu$s for event data | ≤ 300 s for images<br>< 10 $\mu$s for event data |
| Absolute time calibration | 2 $\mu$s, 1 $\mu$s for P/L | < 2 $\mu$s, 1 $\mu$s for P/L |
| Burst Trigger scale | 0.1 s – 100s | 10 ms - 300 s |
| Availability of WFM triggered data | 3 hours | < 3 hours |
| Broadcast of trigger time and position | <30 s after the event for 65% of the events | <25 s after the event for 65% of the events |
| Number of GRB Triggers | Up to 5 GRB triggers per day | > 1 GRB trigger per orbit |

## 3. INSTRUMENT BASELINE DESIGN

The WFM is a coded mask instrument, a concept successfully employed in several X- and gamma-ray instruments on past - e.g., GRANAT/SIGMA, BeppoSAX/WFC - and current - e.g., AGILE/SuperAGILE, INTEGRAL/SPI, INTEGRAL/IBIS, INTEGRAL/JEM-X, Neil Gehrels Swift/BAT - missions. Its working principle is the classical sky encoding by coded masks: the mask shadow recorded by the position-sensitive detectors can be deconvolved to recover the image of the sky, with an angular resolution given by the ratio between the mask element and the mask-detector distance. In order to avoid losing imaging sensitivity, the mask element should not be smaller than twice the detector resolution element.

The position sensitive solid-state detectors of the WFM cameras are the same as those of the LAD - Silicon Drift Detectors (SDDs)[16,25] - but with a modified geometry to get better spatial resolution. SDDs provide accurate positions in one direction but only coarse positional information in the other one; when combined with a 1D coded mask in each WFM camera, they provide "1.5D" positions of celestial sources. Pairs of two orthogonally oriented co-aligned cameras (Fig. 2) give precise 2D source positions (Fig. 3) More details are given in sections 3.2 and 4.1 (optical design of the WFM and SDD detectors description).

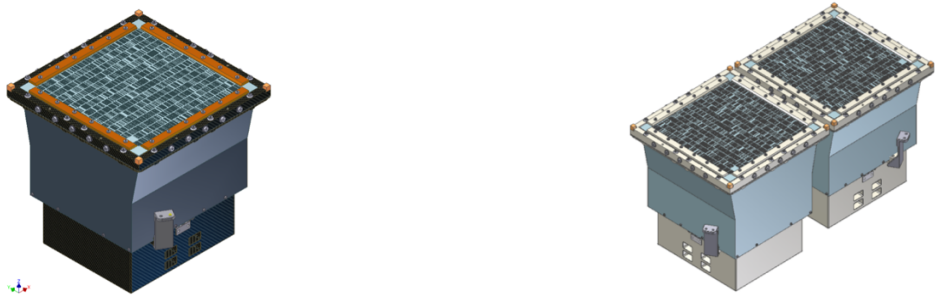

Figure 2. Left: Single WFM camera. Right: WFM camera pair, including two identical cameras arranged orthogonally.

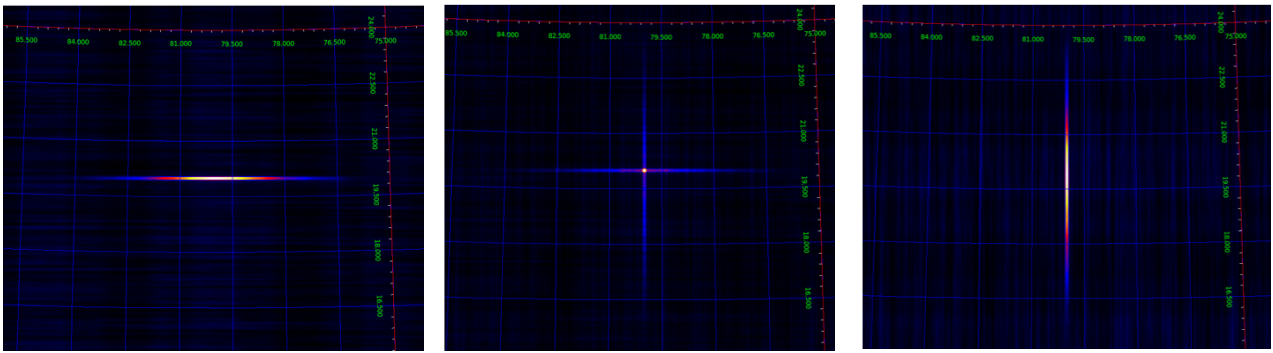

Figure 3. Simulated 2D image of an isolated source (central panel), obtained from the combination of the 1.5D images from the two cameras in a camera pair (left and right panels).

The design of the WFM is fully modular, with six identical cameras grouped in three camera pairs; the two cameras of each camera pair don't need to be placed together. The arrangement of the cameras and camera pairs has been designed to achieve a high level of redundancy. The configuration of the WFM adopted for eXTP (see Fig. 4, left) is inherited from the LOFT/WFM[3] instrument. One camera pair has on-axis direction coinciding with the LAD, SFA and PFA pointing directions, and each one of the two other camera pairs is tilted away ±60º with respect to the on-axis direction. The FoV of each camera (and camera pair) is 90°x90° FoV, at zero response (FWZR), and the fully illuminated FoV is approximately 30°x30° (see more details in section 3.2 and Fig. 7). Therefore, the three camera pairs cover a 180° arc, 180°x90° (FWZR), a large fraction of the accessible sky to the LAD and the other pointed instruments. With this configuration, the WFM can fully support the LAD, SFA and PFA, by providing the triggers of interesting X-ray transients to be observed by them in detail. The WFM has an unprecedented combination of simultaneous FoV - as shown in Fig. 4 (right) - and imaging capability. It achieves the required 1 arcmin source localization accuracy in 2D (see Table 1) and guarantees that any source confusion in the 1º FoV of the LAD - a non-imaging instrument - can be resolved.

The main characteristics of the eXTP WFM instrument are listed in Table 2

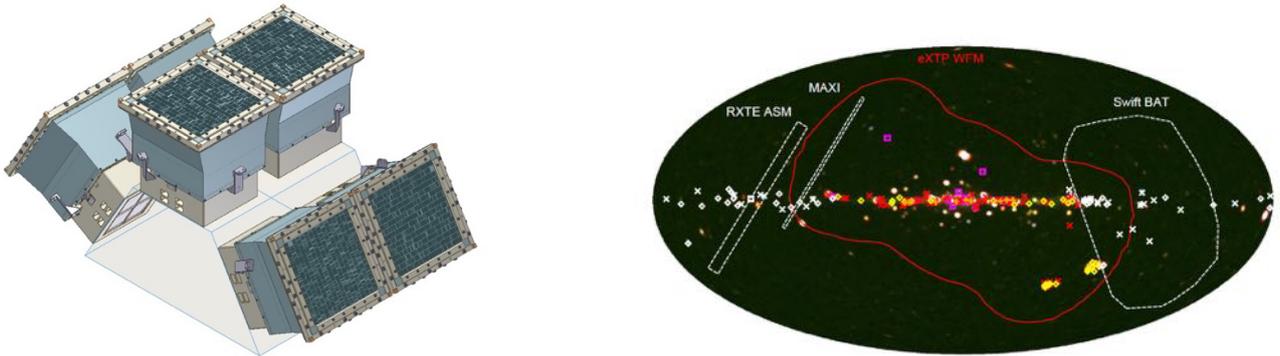

Figure 4. Left: The eXTP/WFM assembly, including six identical coded mask cameras arranged in three camera pairs. Right: The eXTP/WFM FoV, red curve, is compared with other relevant all sky monitoring instruments (see the "Observatory Science with eXTP" paper submitted to Science China, in't Zand et al. 2018). Yellow diamonds mark the known Be X-ray binary systems, red crosses mark supergiant High-Mass X-ray Binaries (HMXBs), orange crosses the Supergiant Fast X-ray Transients (SFXTs), and magenta boxes the Symbiotic X-ray Binaries (SyXBs). Grayed-out sources are those falling outside the WFM FoV during a single pointing toward the Galactic Center. (Background map courtesy of T. Mihara, RIKEN, JAXA, and the MAXI team).

Table 2. Main WFM characteristics

| Item | Value |
| --- | --- |
| WFM cameras<br>ICU | 6 (3 camera pairs)<br>2 units (cold redundancy) |
| Detector type - Number of SDD tiles | Si Drift (SDD) - 24 (4 per camera) |
| SDD spatial resolution (FWHM) | < 60 $\mu$m (fine direction)<br>< 8mm (coarse direction) |
| Detector operating Temperature | - 30ºC < T < - 3ºC |
| Detector effective area | 180 cm$^2$ (per camera) - 360 cm$^2$ (camera pair)<br>1080 cm$^2$ (total: 6 cameras) |
| Energy range | 2 - 50 keV |
| Energy resolution (FWHM) | < 300 eV at 6 keV |
| Mask pitch<br>Size of mask open elements<br>Mask size | 250 $\mu$m x 16.4 mm<br>250 $\mu$m x 14 mm<br>260 x 260 mm$^2$ |
| Mask-detector distance | 202.9 mm |

| Angular resolution | < 5 arcmin (5 arcmin × 5º per camera) |
|---|---|
| Field of view | Camera: 90º × 90º FWZR<br>Camera: 28º × 28º fully illuminated<br>3 Camera Pairs: 180º × 90º FWZR |
| Mass (with margins) | 11.2 kg per camera<br>78.5 kg (total, including ICU) |
| Power (with margins) | 72 W (total, including ICU) |

**3.1 Functional design of the WFM**

The WFM assembly includes three camera pairs and the instrument control unit (ICU, two units in cold redundancy). Each camera has a detector tray, with four detector assemblies (DAs), four Beryllium windows to protect the detectors, one back end electronics (BEE) including a power supply unit (PSU), a coded mask (with mask frames and a thermal blanket) and a collimator (see Fig. 5). The DA consists of a detector (SDD) and the corresponding front end electronics (FEE), as described in more detail in sections 4.2 and 5.2

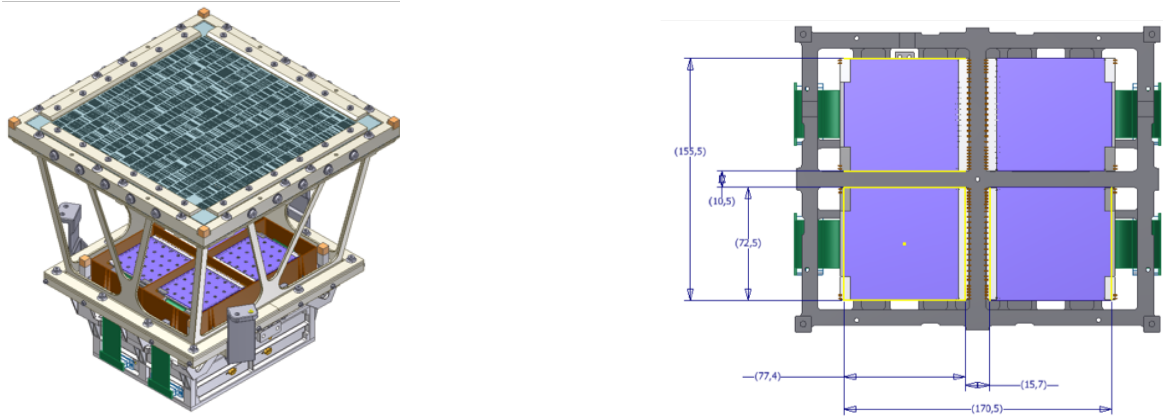

Figure 5. Left: Scheme of a WFM camera, consisting of the coded mask assembly, the collimator (without walls in this figure), the detector plane assembly and the back end electronics (bottom). Right: scheme with dimensions of the detector plane, consisting of four detector assemblies.

The functional block diagram of the WFM is shown in Fig. 6. The main functionalities of the different subsystems are:

- WFM FEE: forward filtered bias voltages to the SDDs, provide power and configuration data to the ASICs, read-out and analogical/digital (A/D) convert the SDD signals, interface the BEE, provide mechanical support to the SDD.

- WFM BEE: time tagging of the X-ray events, trigger filtering, pedestal subtraction, common mode noise subtraction, determination of charge cloud center and width (position in the fine and coarse direction), determination of the total charge collected - correcting for channel gains - to get the energy, event packet generation.

- The ICU hosts the mass memory, the power distribution unit (PDU) and the data dandling unit (DHU):
    - DHU of the WFM ICU: interfacing the BEEs to the on-board data handling unit (OBDH), telecommand execution (TC), instrument monitoring and configuration, on board time management, image integration, burst triggering.

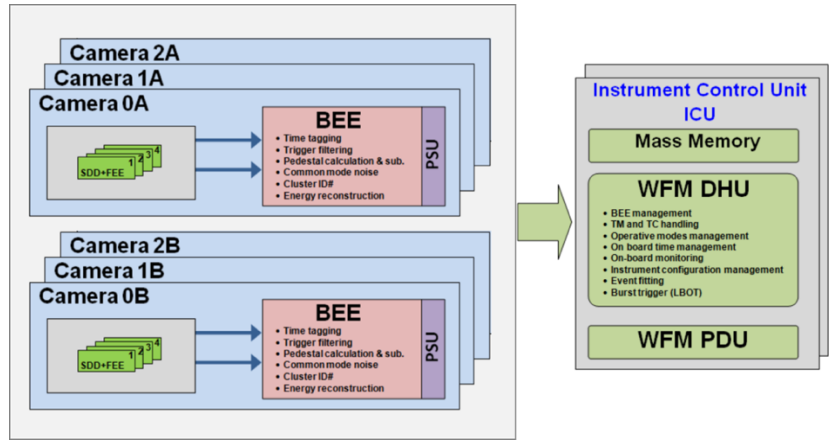

Figure 6. Functional block diagram of the WFM, made of three camera pairs with a total of six identical cameras, plus two instrument control units (in cold redundancy). For each camera there are four DAs (SDDs plus their corresponding FEEs), and one BEE module which includes the power supply unit (PSU). The ICU (two in cold redundancy) include mass memory, power distribution unit (PDU) and data handling unit (DHU).

## 3.2 Optical design

The geometry of the WFM cameras is shown in Fig. 7, left (and Fig. 5, right). It is driven by the asymmetric response of the detectors, with a coarse resolution of 8 mm and a fine resolution of 60 $\mu$m (see detailed explanation in section 4.1 about detectors). The coded mask has been designed accordingly (Fig. 7). It is made of 150 $\mu$m thick Tungsten (good for the energy range of the SDDs), and its pattern consists of 1040x16 open/closed elements. The mask pitch is 250 $\mu$m x 16.4 mm: dimensions of the open elements are 250 $\mu$m x 14 mm and spacing between the elements in the coarse direction is 2.4 mm. A 25% open mass fraction has been chosen, to improve sensitivity to weaker sources, but this is open to trade-off. The detector-mask distance is 202.9 mm (see Fig. 7). The angular resolution (FWHM) for the on-axis viewing direction is the ratio of the mask pitch and the detector-mask distance, 4.24 arcmin in the fine resolution direction and 4.6º in the coarse resolution direction. More details about the optical design of the WFM are listed in Table 2.

To give accurate source positions, the mask must be flat (or at least maintain its shape) with a tolerance of ±50 $\mu$m over its entire surface across the full operational temperature range. A Sun shield is needed to avoid uncontrolled deformations of the coded mask. Another essential factor to fulfill this requirement is the design of mask frame, as for instance, the mask frames and the pretension mechanism. More details about the WFM mechanical and thermal designs, aimed to guarantee the required spatial resolution of the WFM, are given in section 5.

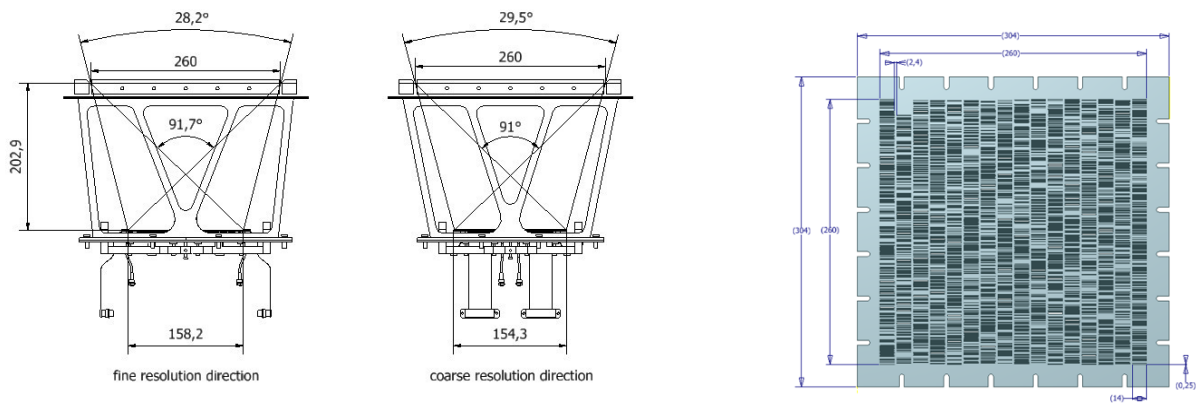

Figure 7. Left: Optical configuration of a WFM camera showing the size of the mask and the detector plane and the distance between them (in mm). The angles for the fully illuminated and zero response are also shown (28º and 90º approximately). Right: coded mask design with dimensions.

# 4. WFM DETECTORS, ELECTRONICS AND INSTRUMENT CONTROL UNIT

The electrical architecture of the eXTP WFM will be similar to that of the LOFT/WFM[5]. It encompasses all electrical subsystems of the WFM and their interfaces with the spacecraft platform. A scheme taken from the LOFT design adapted to the six cameras of eXTP is shown in Fig. 8.

The central part of the WFM, from the electrical point of view, is the ICU. It includes the DHU, mass memory and the PDU (see also Fig. 6). There is a main and a redundant ICU unit, housed in two separate boxes. A burst on-board trigger functionality is implemented as a part of the DHU. The DHU is directly attached to the eXTP spacecraft OBDH system. The electrical interface is assumed to be SpaceWire. Mass memory stores the WFM data until downlink, and each BEE is connected to both the main and the redundant DHU. The bus power is routed through the WFM PDU providing ON/OFF switching.

The detectors, FEE and BEE for the LAD and the WFM share almost identical requirements on their performance and functionality. Therefore, a common approach for both instruments has been chosen, making use of similar detector design, ASIC and FPGA devices in order to reduce complexity, facilitate the development as well as the later calibration and operation of the two instruments.

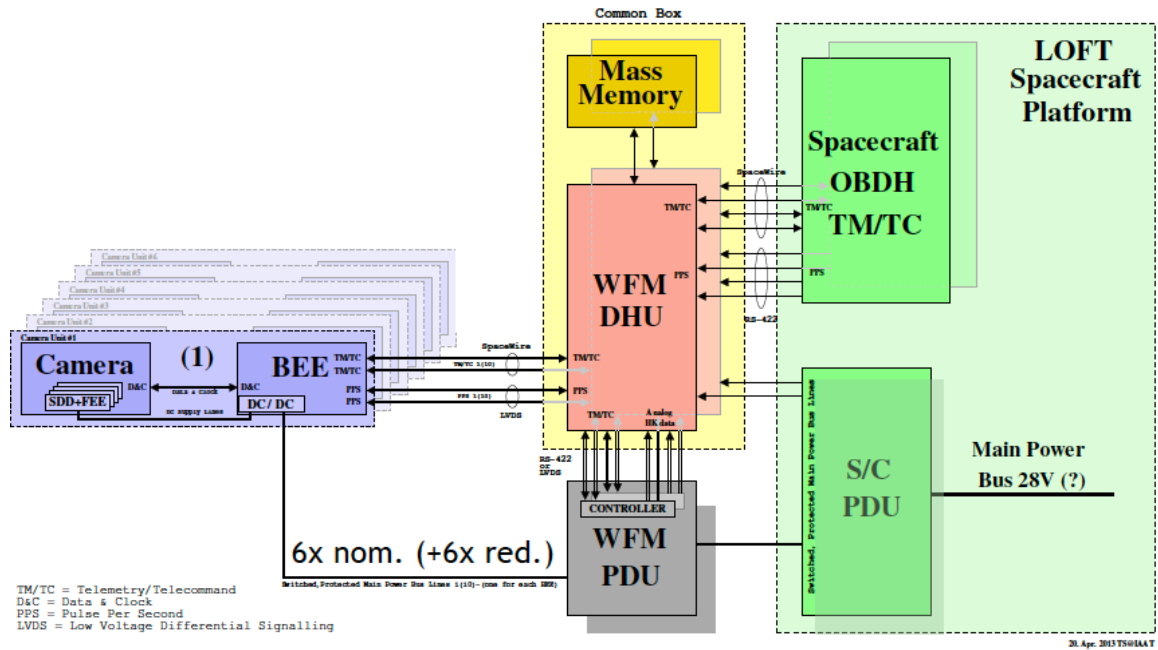

Figure 8. The electrical architecture of the eXTP WFM (six cameras), adapted from LOFT.

## 4.1 The WFM Silicon Drift Detector

The WFM detector plane is based on the same large area SDD technology developed for the LAD (heritage from the ALICE experiment at the Large Hadron Collider (LHC) at CERN)[20,21] but optimized to enable 2D spatial resolution imaging. When a photon is absorbed by the SDD, it generates an electron cloud that is focused on the middle plane of the detector, and then drifts towards the anodes at constant speed (see Fig. 9, left). During the drift time, the electron cloud size increases due to diffusion. The cloud that arrives at the anodes can be described as a Gaussian, that has an area equal to the total charge (i.e., the photon energy), a mean value representing the "anodic" coordinate of the impact point and a size σ which depends on the "drift" coordinate of the absorption point. At a distance d from the photon absorption point, the size of the Gaussian-shaped cloud is given by the following equation:

$$\sigma \sim \sqrt{\frac{2\,k_B T}{qE} d}$$

where $k_b$ is the Boltzmann constant, T is the temperature, q is the electron charge and E is the drift electric field (Fig. 9, left). The analysis of the charge distribution over the anodes is performed on-board by the FPGA-based BEE on each event and therefore results in the determination of the amplitude, the anode position and the drift position of the event.

In order to maximize the signal-to-noise ratio of the anode and drift position information, the design of the WFM SDD has been optimized by means of Monte Carlo simulations[10]. This results in a smaller anode pitch than for the LAD one. The current values for the eXTP/WFM SDDs are: anode pitch 169 $\mu$m (versus 970 $\mu$m for the LAD), drift length 35 mm and quasi-squared overall dimensions (7.74 x 7.25 cm² geometric area, 6.50 x 7.00 cm² effective area, see Fig. 9 left). Si thickness is 450 $\mu$m. Taking into account the eXTP radiation and thermal environments, the relevant parameters for the detector and read-out electronics and the current choice of the anode pitch the expected performance in terms of energy resolution and position are shown in Figs. 9 (right) and 10, respectively: energy resolution is better than 300 eV at 6 keV; position resolution is better than 60 $\mu$m in the fine direction and 8 mm in the coarse one (see also Table 2)

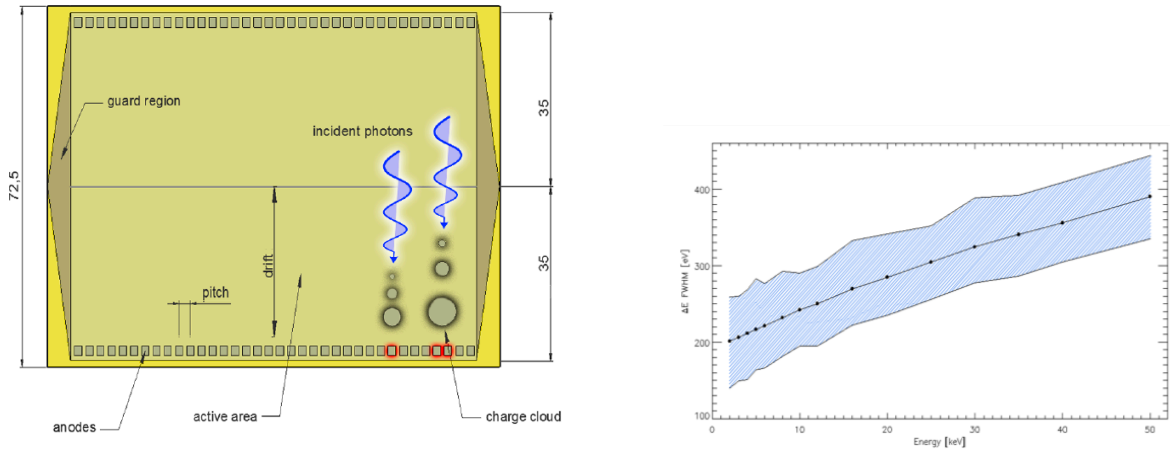

Figure 9: Left: working principle of the SDD detectors and dimensions, in mm, of the SDD tile (geometric width: 77.4 mm). Right: Energy resolution (FWHM) as a function of the photon energy, for the WFM detectors as a function of the photon energy. The shaded area shows the range of energy resolution for each photon energy, whereas filled circles show the resolution values averaged on the whole SDD channel.

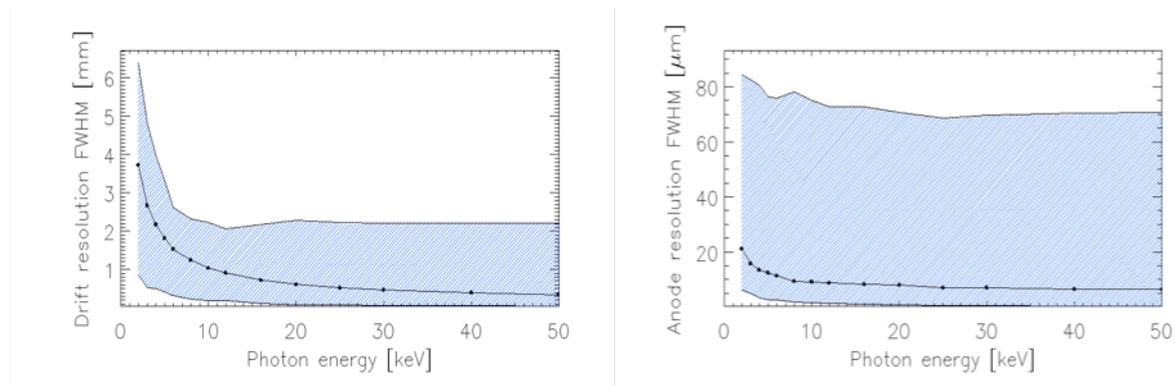

Figure 10. Drift (left) and anodic (right) spatial resolution (FWHM) as a function of the photon energy, for the WFM detectors (SDDs). The shaded area shows the minimum-maximum spatial resolution performance inside the detector channel. Filled circles show the averaged channel values.

## 4.2 Front and Back End Electronics

The front-end electronics design consists of several ASICs that are directly located on the backside of the DA (a Printed Circuit Board (PCB) with an SDD/FEE "sandwich"). The required number of ASICs per SDD is higher than for the LAD: 24 x IDeF-X HD ASICS[17], with a smaller pitch than LAD ones (as said, 169 $\mu$m versus 970 $\mu$m), and 2 x OBW-1 ASICS[1].

The number of read-out anodes per tile is: 12 ASICS x 32 channels/ASIC x 2 rows = 384 x 2 rows = 768 total (to be compared to 896 total for LOFT/WFM). The dimensions of the Si tile have not changed significantly.

When the collected signal charge exceeds a programmable threshold in one of the ASIC channels, a trigger is generated and forwarded to the BEE. In case of a confirmed valid trigger pattern, the collected signals in all ASICs of the respective detector half are then digitized and passed on to the BEEs.

Following the A/D conversion, the BEE event processing pipeline will be activated, where a time tag will be added to each event and a pedestal and common noise subtraction will be performed. In addition, an energy reconstruction takes place to determine the event parameters. Besides the event processing, the BEE controls the FEE and PSU operation, generates housekeeping data and rate meters, and transmits formatted event packets onward.

For the WFM, there is only one level of BEE, as there are only four detectors to be read out per camera. One BEE is located along with the PSU at the bottom of each camera. There is a higher level of processing power required inside the BEEs of the WFM (besides the standard processing pipeline that is common with the LAD) because of the need to calculate the position of each photon in the detector plane accurately from the FEE data. This allows to integrate high-resolution detector images in several energy bands directly on board. For this purpose, the WFM BEE will be based on an RTG4 FPGA that is generally larger, faster and more flexible than the RTAX-SL. As in the Panel BEE of the LAD, the BEE of the WFM will transfer the science data products to the WFM ICU along with the House Keeping (HK) data via a Space Wire interface.

### 4.3 Instrument Control Unit

The Instrument Control Unit forms the central controlling element of both the instruments. It provides the interface to the spacecraft On Board Data Handling (OBDH) and also access to all instrument sub-systems via Space Wire. Each ICU box consists of three components that are identical for both instruments: Data Handling Unit, mass memory and Power Distribution Unit (see Fig. 8). The WFM ICU in addition has an FPGA board to perform on-board computations necessary to locate bright transient events in the images in real time. The standard tasks performed at the level of the ICU involve telecommand execution and distribution, access to mass memory, time distribution and synchronization, data processing and compression, HK collection, instrument health monitoring and calibration tasks. The ICU boxes of both instruments contain each PCB board twice for cold redundancy.

## 5. WFM MECHANICAL AND THERMAL DESIGN

The mechanical and thermal design of the WFM camera must achieve stable and accurate imaging properties of the coded mask - detector plane system. Therefore, it is important to assure a stable parallelism between the coded mask plane and the detector plane. The main elements of the camera are depicted in Fig.11, left.

### 5.1 Coded mask, masks frames and collimator

The coded mask is manufactured from a 150 $\mu$m thick Tungsten foil and has a coded area of 260 mm x 260 mm (see Fig. 7). The choice of Tungsten among the high atomic number materials is justified by the heritage of SuperAGILE. The assembly of the coded mask is performed by means of top and bottom frames, as shown in Fig. 11, right. The mask frame set acts as a pretension mechanism in order to minimize the vertical displacements of the mask during the operational mode, and thus fulfill the mask flatness requirement. A Sun shield is also required, to accomplish the mask temperature gradient requirement (see details in the optical design section 3.2). Detailed description of the camera thermal design is found in section 5.3.

There are strict requirements on the flatness and stability of the coded mask. For instance, the mask must be flat (or at least maintain its shape) to ±50 μm over its entire surface across the full operational temperature range. This requirement implies that we cannot tolerate large gradients in temperature and consequently that the temperature excursion between the sunlit and the dark parts of the orbit must be less than 10°C. This is the prime reason why a Sun shield is a requirement for the WFM; another essential factor to fulfill the mentioned requirement is the design of mask frame.

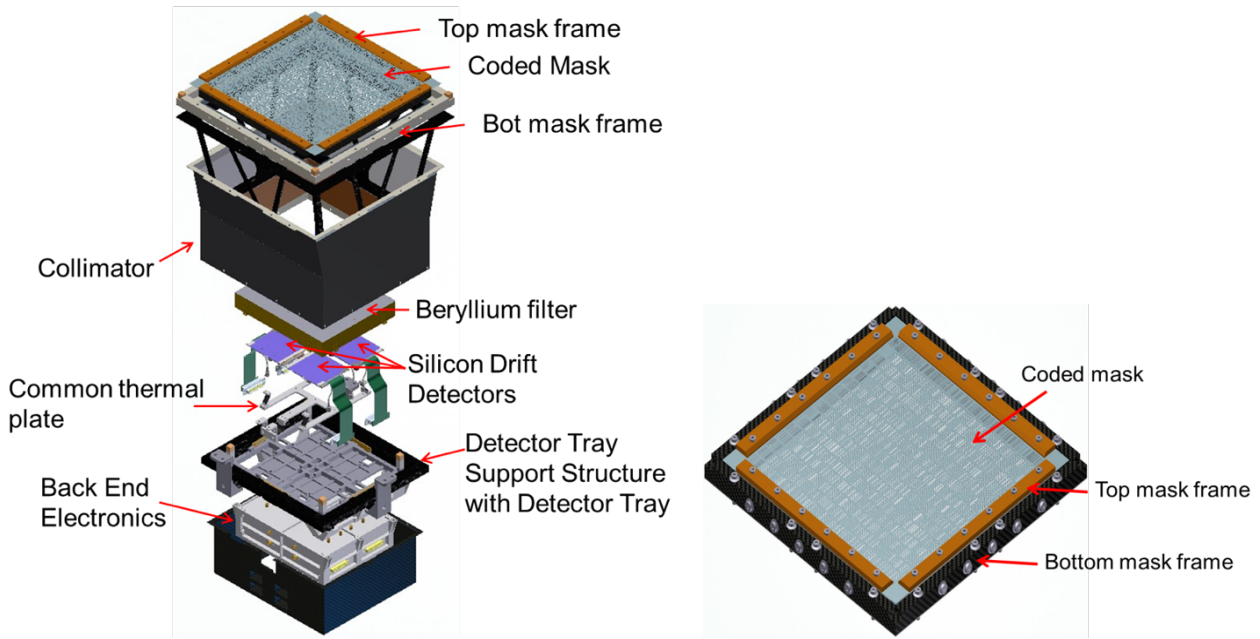

Figure 11. Left: Exploded view of a single WFM camera. Right: Coded mask assembly, showing the mask frames.

The collimator supports the coded mask assembly; it is made of an open CFRP structure, 3 mm thick, as depicted in Fig. 12, left. This structure has enough stiffness to avoid deformations that can appear during launch (accelerations) and operation (thermal stresses). The outer part of the collimator, made of 1 mm thick CFRP layers, will be covered by a 150 $\mu$m thick Tungsten sheet, acting as a background shield. On the other hand, Copper and Molybdenum plates - 50 $\mu$m thick - will be placed in the inner part of the collimator, for in-flight calibration purposes, as shown in Fig. 12, right.

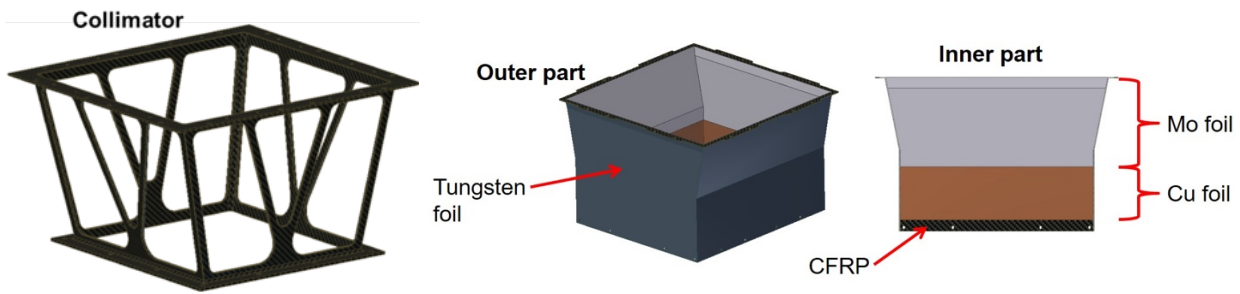

Figure 12. Left: Collimator structure. Middle: Collimator outer shielding made of Tungsten. Right: Cross-section of the collimator inner shielding.

### 5.2 Detector box

The detector box consists of three elements: the detector tray, the detector tray support structure (DTSS) and the BEE box, as shown in Fig. 13 (see as well Fig. 11). The detector tray includes the detector support plate (DSP) and four detector assemblies. In each DA, the SDD tile is glued on a ceramic PCB that contains the FEE; 24 ASICs are wirebonded to read-out all the detector anodes as shown in Fig. 14 right. The FEE PCB is mechanically supported by an Invar bracket. There is an Aluminum thermal plate with a thermal strap between the DA and the Invar bracket, to remove the heat generated by the DAs (Fig. 14, left). Three Invar mounting legs (highlighted in red in Fig. 14, left) provide the mechanical interface to the DSP.

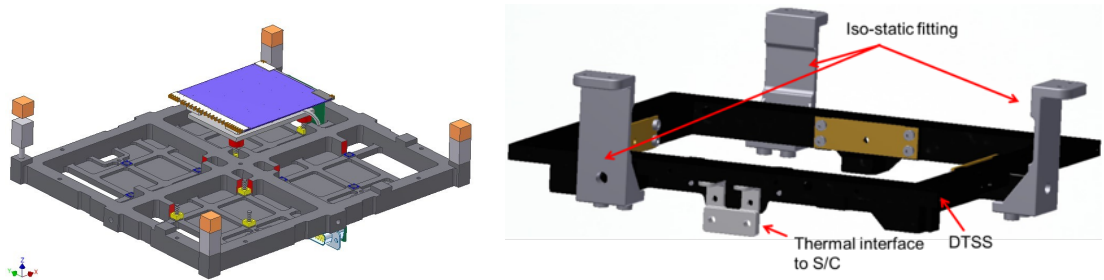

Figure 13. Left: The detector tray assembly with the detector support plate and one detector assembly (DA). Right: Detector tray support structure.

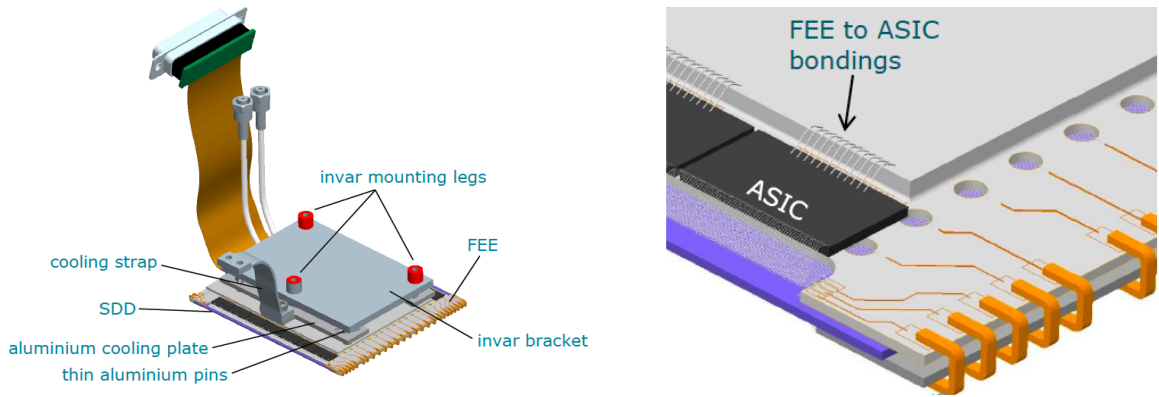

Figure 14. Left: WFM Detector Assembly, showing the High and Medium Voltage cables (HV/MV) from BEE (this is an upside-down view). The invar mounting legs are the mechanical interface for alignment of DA with DSP. Right: Detail of the FEE with ASICS, showing the bondings from SDD anodes to ASIC inputs and between ASICs and the FEE board.

The DSP provides a referenced allocation for the four DAs, to guarantee good mechanical alignment in the whole detector plane. The DSP is made of Invar in order to match with the coefficient of thermal expansion (CTE) of the detector assembly. On the other hand, the DTSS holds the detector support plate and facilitates the mounting of the collimator and the BEE box. In addition, the iso-static fittings (mounting legs) are fixed on the DTSS itself and act as the mechanical interface of the WFM Camera with the spacecraft. The design of the mounting legs has been updated to be compatible with the eXTP spacecraft requirements.

A Beryllium filter is included above the detector plane. Si drift detectors will suffer from the impacts of micro-meteorites and small particles of orbital debris due to the large FoV of the WFM camera[19]. This effect can be mitigated with the placement of a 25 μm thick Beryllium filter on a fiberglass structure ~ 8mm above the detector plane to prevent HV discharges, as depicted in Figure 11. The fiberglass structure will be placed on the DSP, forming four isolated and light tight compartments for SDDs which will not obstruct their field of view. The big advantage of using four isolated compartments is that in case the Beryllium filter is damaged (by debris), only one SDD will be lost instead of all. With the Beryllium filter in combination with the thermal blanket in front of the coded mask, the risk of impact of particles of size ~100 μm is reduced to about $1.6 \times 10^{-2}$ per year per camera[12,19].

### 5.3 Thermal design of the WFM cameras

The thermal design of the camera should minimize variations of temperature across the coded mask and set a stable temperature on the four detector assemblies. First, direct Sun illumination of the mask should be avoided and therefore a Sun shade or alike should be provided by the spacecraft. In addition, the coded mask is covered by a thermal foil which has an acceptable transparency down to the softer X-rays, with energy 2 keV. The thermal foil chosen as baseline is Sheldahl 146545, with three layers, from outside to inside 0.15 μm SiO2, 100 nm VDA and 7.6 μm Kapton. It will help to reduce the mask temperature gradients along the orbit and across the mask surface. Second, the operating temperature of the four detector assemblies is in the range between -30ºC and -3ºC. In order to fulfill this thermal requirement, the cooling plates of all four detector assemblies (see Figure 14, left) are connected to the common thermal plate via thermal straps in order to dissipate the internal heat from the detector assemblies. The material chosen for the common thermal plate is

Aluminum because it has high thermal conductivity and low weight. The common thermal plate should be connected to the thermal interface (I/F) provided by the spacecraft (S/C) to get the optimum operation temperature of the detectors. In addition, an MLI blanket will wrap the collimator in order to provide the detector assemblies with optimal thermal insulation, and to keep the camera alignment in the desired margin during the mission.

The power dissipation of the back-end electronics is expected to be radiatively transferred to the optical bench and deep sky by means of the backside of the box itself, i.e., that side will act as a radiator.

A Detailed Geometrical and Thermal Mathematical Model (DGMM/DTMM) has been defined for a generic camera pair. The model includes 73 nodes which represent the camera elements and five external nodes (deep sky, Optical Bench, sunshield and I/F of S/C) as shown in Table 3. The thermal model has helped to verify that the thermal requirements are accomplished. For instance, the mask surface has been modelled with 9 nodes in order to assess temperature gradient across the mask is less than 5 ºC as required. The thermal model is based on the mechanical design (Figure 11). This fact is illustrated in Figure 15 where the bulk properties of the cameras elements are represented.

Table 3. Thermal nodes description of the DTMM camera

| | Item | Nodes |
|---|---|---|
| WFM camera | Mask | 18 nodes (9 outer, 9 inner) |
| | Top Mask Frame | 8 nodes (4 outer, 4 inner) |
| | Bottom Mask Frame | 8 nodes (4 outer, 4 inner) |
| | Collimator | 8 nodes (4 outer, 4 inner) |
| | SDD | 8 nodes (4 outer, 4 inner) |
| | FEE | 8 nodes (4 outer, 4 inner) |
| | DTSS | 2 nodes (1 outer, 1 inner) |
| | BEE box | 12 nodes (6 outer, 6 inner) |
| | Thermal Common Plate | 1 node |
| S/C | OB | 1 node |
| | Sunshade | 2 nodes (1 outer, 1 inner) |
| | I/F of S/C (TBC) | 1 node |

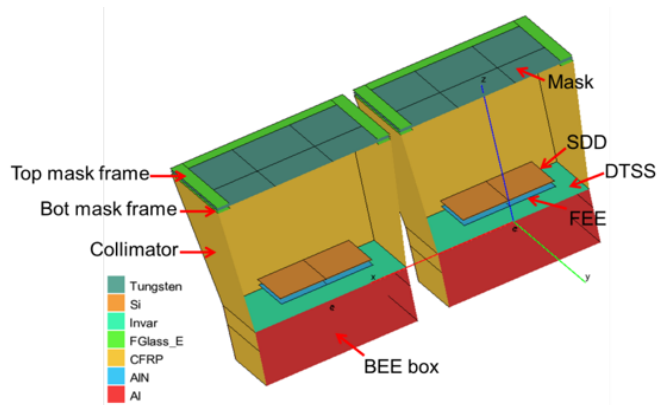

Figure 15. Bulk properties of the elements within the WFM camera thermal model.

# 6. CONCLUSIONS

The Wide Field Monitor instrument as first designed for LOFT offers a modular design, with a number of camera pairs that has been adjusted to fulfill the science goals and resources of the eXTP mission. The accommodation in the spacecraft is flexible, as long as the field of view is unobstructed and the cameras are protected from direct Sun illumination.

The unprecedented combination of large field of view, and sensitivity – both imaging and spectral - makes it a crucial instrument not only as the provider of the triggers for the eXTP pointed instruments – SFA, LAD and PFA – but also as an excellent instrument to do its own science.

# ACKNOWLEDGMENTS

The Spanish team acknowledges support from the MINECO grant ESP2017-82674-R and FEDER funds. The Chinese team acknowledges the support of the Chinese Academy of Sciences through the Strategic Priority Research Program of the Chinese Academy of Sciences, Grant No. XDA15020100. The SDD development results described in this paper were obtained within the Italian INFN project ReDSoX2 (Research Detectors for Soft X-rays), also supported under ASI agreement 2016-18-H.0, INAF grant TECNO-INAF-2014, FBK-INFN agreement 2015-03-06. The German team acknowledges support from the Deutsche Zentrum für Luft- und Raumfahrt, the German Aerospce Center (DLR). The Polish Team acknowledges the support of Science Centre grant 2013/10/M/ST9/00729.